\documentclass[11pt]{article}
\usepackage[margin=1in]{geometry}
\usepackage{graphicx}
\usepackage{url}

\def\beq{\begin{equation}}
\def\eeq#1{\label{#1}\end{equation}}
\def\eeqn{\end{equation}}

\def\beqa{\begin{eqnarray}}
\def\eeqa#1{\label{#1}\end{eqnarray}}
\def\eeqan{\end{eqnarray}}

\let\bar=\overbar

\def\Dslash{\not{\hbox{\kern-4pt $D$}}}
\def\dslash{\not{\hbox{\kern-2pt $\del$}}}

\def\msb{{\bar{\ssstyle M \kern -1pt S}}}

\def\Title#1{\begin{center} {\Large {\bf #1} } \end{center}}
\def\Author#1{\begin{center} {\normalsize {\sc #1} } \end{center}}
\def\Institution#1{\begin{center} {\normalsize {\it #1} } \end{center}}
\def\Abstract#1{\noindent {\normalsize {\bf Abstract:} {\normalfont #1}}}
\def\Conference{\vspace{4mm}\begin{raggedright} {\normalsize {\it Talk presented at the 2019 Meeting of the Division of Particles and Fields of the American Physical Society (DPF2019), July 29--August 2, 2019, Northeastern University, Boston, C1907293.} } \end{raggedright}\vspace{4mm}}

\begin{document}

%
%

\Title{Acts: A common tracking software}

\Author{Xiaocong Ai for the ACTS developers}

\Institution{Physics Department, University of California, Berkeley CA, United States of America}

\Abstract{The reconstruction of charged particle trajectories is one of the most complex and CPU consuming parts of event processing in high energy experiments. At future hadron colliders such as the High-Luminosity Large Hadron Collider (HL-LHC) or the Future Circular Collider (FCC), the significantly increased number of simultaneous collisions will result in a much more challenging tracking environment. Concurrent algorithms exploiting modern computing architectures with many cores and accelerators are necessary to maintain and improve the tracking performance. Based on the tracking experience at LHC, the ACTS project is an attempt to encapsulate the current ATLAS software into a experiment-independent and framework-independent software designed for modern computing architectures. It provides a set of high-level track reconstruction tools which are agnostic to the details of the detection technologies and magnetic field configuration. The software has been fully tested for thread-safety to support parallel execution of the code and its data structures are optimized for vector operations to speed up linear algebra operations.}

\Conference

%
%

\section{Introduction}

The record-breaking data taking of the LHC in the second run (Run-2) between 2015 and 2018 provides chance for extraordinary exploration of the high-energy frontier. To greatly increase the sensitivity to new physics, the LHC will enter the HL-LHC~\cite{HL-LHC} era in 2026 with an instantaneous luminosity up to $L = 7.5 \times 10^{34} \texttt{cm}^{-2} \texttt{s}^{-1}$, which corresponds to approximately 200 inelastic proton-proton collisions per beam crossing (pile-up).
The greatly increased number of concurrent tracks from pile-up will put great pressure on CPU consumption at the LHC experiments.  
Highly-performant tracking software with concurrent algorithms exploiting modern computing architectures with many cores and accelerators have to be developed. 

The A Common Tracking Sofware (ACTS) is an attempt to prepare a tracking toolkit for modern computing architectures and future colliders based on the ATLAS~\cite{ATLAS} tracking experience with long term maintenance in mind. 
While the current ATLAS tracking software~\cite{atlas-track-sw} within the ATLAS software framework (Athena)~\cite{athena} has shown good overall performance at the LHC Run-1 and Run-2, many tracking components were not developed with a multi-threaded operation mode in mind during their design phase. The migration of Athena to multi-threaded Athena (AthenaMT)~\cite{athenaMT} to adapt to the multi-core environment is on-going, which requires rewriting a significant amount of code in Athena. This makes the ACTS a possible alternative to the migration. It puts special emphasis on thread-safety in order to support an multi-threaded event processing and is designed to be independent of any event processing framework. ACTS also serves as on open-source platform for algorithm development for future track reconstruction.

With the advent of continuous integration (CI) techniques and git based code development, more emphasis can be put on the compilation state of the code. Hence, ACTS tries to minimise virtual interfaces and on the contrary enhances the use of compiler templating. The price of longer compilation time is hereby outweight by generally fast code execution. Following this design principles, ACTS uses C++ concept mechanisms rather than virtual interfaces in order to define code structures and modules.
 
ACTS is divided into two main components \texttt{acts-core}~\cite{acts-core} and \texttt{acts-framework}~\cite{acts-framework}. The \texttt{acts-core} contains the detector-independent tracking toolkit. The \texttt{acts-framework} is a \texttt{Gaudi}~\cite{gaudi} inspired test framework, which can be used to run the continuous integration tests to ensure quality and thread-safety of the code. It allows a fast development and testing turn-around time, which is more flexible than the large software stacks of the LHC experiments. In addition to \texttt{acts-core} and \texttt{acts-framework}, the ACTS also includes a fast simulation component \texttt{acts-fatras}~\cite{acts-fatras} to simulate the trajectories of particles in detector with simplified material effects. For instance, input datasets used by various tracking algorithm development projects such as the Kaggle TrackML challenge~\cite{trackML} and the HEP.TrkX project~\cite{trkx} are provided by \texttt{acts-fatras} simulation with a prototype detector,~i.e. the TrackML detector~\cite{trackML}.

\section{ACTS tracking components}
At the time of writing, the current release of the \texttt{acts-core} is v0.10.04~\cite{acts-core-tag}. The basic components needed for track reconstruction such as tracking geometry, tracking Event Data Model (EDM) and propagation engine are well-developed. Prototypes of algorithmic tracking tools including seed finding, track fitting and vertex reconstruction are implemented. The interface to support concurrent track reconstruction with multiple alignment constants, calibration constants or even magnetic field has been developed. 

The detector geometry description is necessary for track propagation and material effects integration during track reconstruction. Since track reconstruction with an accurate detector description as used in full detector simulation could require large CPU time consumption, an abstraction of the detector geometry,~i.e. tracking geometry, is used in track reconstruction in ATLAS tracking software. In the ACTS, the same concept of \texttt{TrackingGeometry} is used for track reconstruction. The \texttt{Surface} is the most fundamental geometrical object and could be extended to \texttt{Layer}s and  \texttt{Volume}s. The ACTS \texttt{TrackingGeometry} can be built from various geometry description such as the ATLAS \texttt{GeoModel}~\cite{geomodel} description and the \texttt{DD4Hep}~\cite{dd4hep} modelling. An abstract \texttt{DetectorElementBase} class will help rebuild the connection between the full detector geometry description and the tracking geometry. ACTS \texttt{TrackingGeometry} supports various sub-detector such as the Silicon tracker, Calorimeter and Muon Chambers. 

The trajectory of a charged particle in magnetic field can be described with a set of track parameters such as space coordinates and the momentum at that space point. ACTS also includes an additional time parameter per track to support timing detectors. Inheriting from the base class \texttt{TrackParameters} with the parameter set $(loc1, loc2, \phi, \theta, \frac{q}{p}, t)$, the \texttt{SingleBoundParameters} and \texttt{SingleCurvilinearParameters} are used to describe a single component trajectory in the local frame of a detector surface and in the curvilinear frame, respectively. To support the multi-component fitter such as Gaussian-Sum-Filter (GSF)~\cite{gsf} which describes the non-Gaussian energy loss of electron as a weighted sum of several Gaussian distributions, a multi-component trajectory is described by \texttt{MultivariantBoundParameters} and \texttt{MultivariantCurvilinearParameters}. 

To propagate initial track parameters throughout the detector, a highly-flexible  \texttt{Propagator} is developed with the main task of integrating the motion of the particle governed by the Lorentz force in a magnetic field to the transport of track parameters. The  \texttt{Propagator} is designed to be templated on a \texttt{Stepper} and a \texttt{Navigator} to support user-defined integration of particle motions and the tracking geometry. 
The adaptive Runge-Kutta-Nyström method~\cite{stepper} is used as the primary integration method in ACTS.
The \texttt{propagate} call of the  \texttt{Propagator} also has a highly-templated design to allow for propagation with various EDMs for track parameters and surfaces, and user-defined options including the execution of a list of actions,~i.e. \texttt{Actors} at each integration step, and abort conditions  \texttt{Aborters}. Material effects could be included by adding a  \texttt{MaterialInteractor} in the  \texttt{Actor}. 

\subsection{Track finding and fitting}
After the detector-dependent measurement objects such as clusters or drift circles are formed from the detector response, track reconstruction uses local or global pattern recognition algorithms to identify the group of measurement objects that stem from the same particle. The local pattern recognition method usually starts with the seed finding process to search for combinations of measurement objects (typically doublets or triplets) originating from the same particle trajectory and provide an initial estimation of direction of the trajectory. Starting from the reconstructed seeds, a combinatorial Kalman filter technique is used to build all trajectory candidates in parallel by progressively adding other measurements along track propagation. In this approach, track fitting is integrated with track finding in the senset that track parameters are updated using Kalman filter technique during the propagation. 

Inspired by the seed finder in ATLAS tracking software, ACTS implements a detector-independent \texttt{Seedfinder} to build seeds from triplets of \texttt{SpacePoint}s (created from either a single two-dimensional measurement, or a combinatation of two one-dimensional measurements): starting from a (middle) \texttt{SpacePoint}, \texttt{SpacePoint} closer (inner) or further (outer) to the \texttt{SpacePoint} are searched for within a window in the azimuthal angle. The inner \texttt{SpacePoint}s and outer \texttt{SpacePoint}s are then checked with the middle \texttt{SpacePoint} to form a triplet by comparing their polar angle. The selection for both the inner (outer) \texttt{SpacePoint} and the triplet are motivated by the helix model of track trajectory in an assumed homogeneous magnetic field: a circle in the transverse plane and a linear trajectory along the beam direction. Highly configurable selection criteria are applified to ensure both the purity and efficiency of the seeds by taking care of various properties such as potential minimum transverse momentum of the tracks, measurement errors, and region of interest for seed finding.

\begin{figure}[htbp]
\centering
\includegraphics[width=0.46\columnwidth]{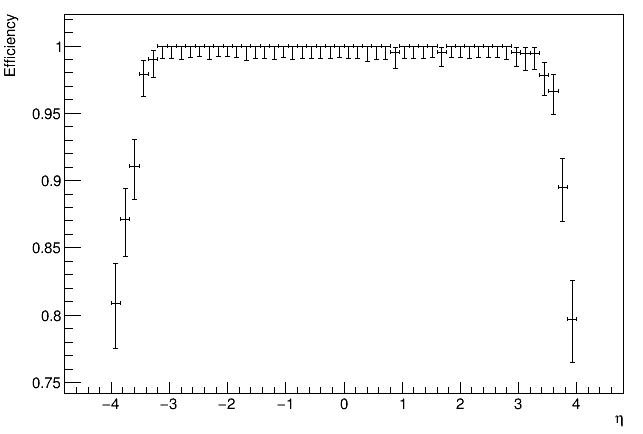}
\includegraphics[width=0.46\columnwidth]{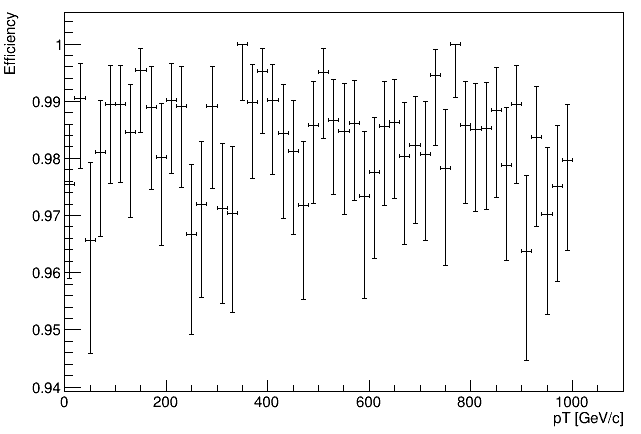}
\caption{ACTS \texttt{KalmanFilter} efficiency as a function of eta (left) and pT (right) for a sample of 10000 muons generated with \texttt{acts-fatras} with the TrackML detector.}
\label{fig:trackeff}
\end{figure}

\begin{figure}[htb]
\centering
\includegraphics[width=1\columnwidth]{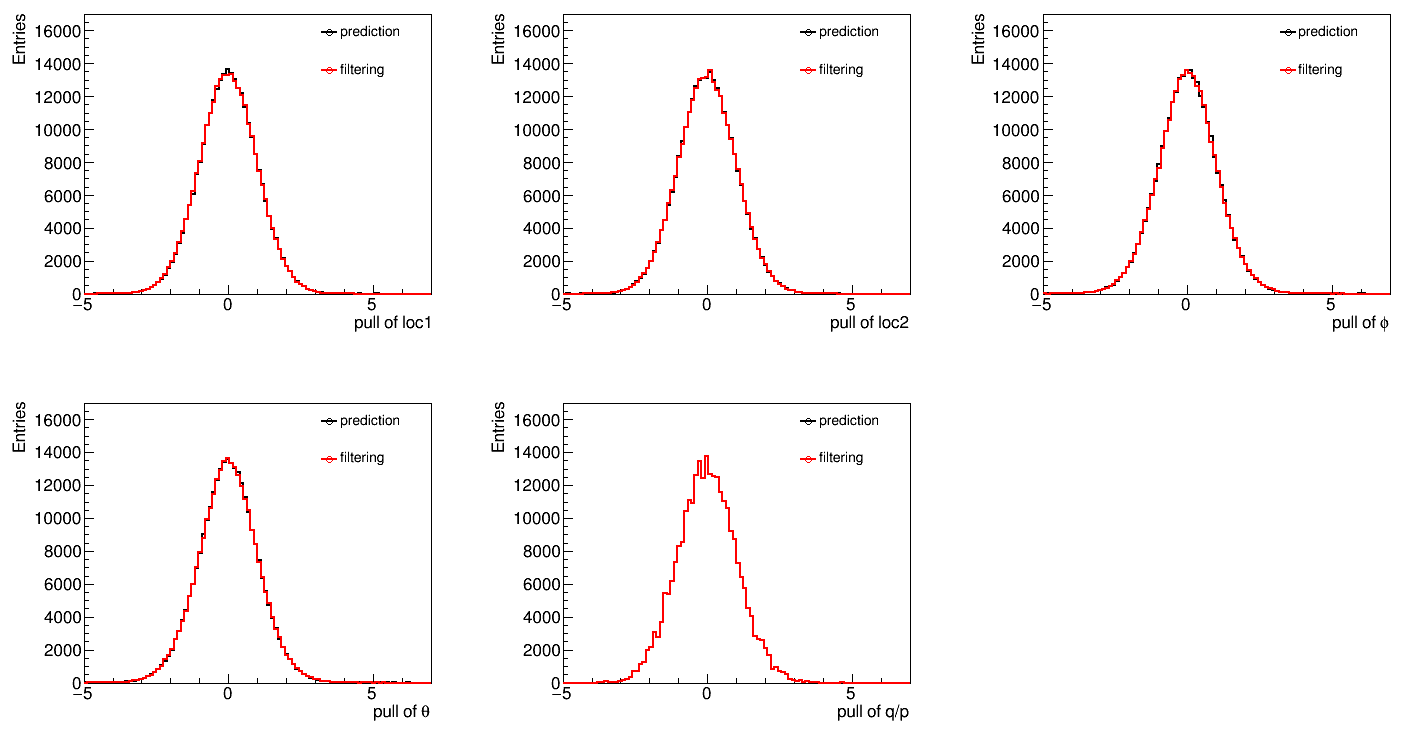}
\caption{Distributions of pull values of the track parameters for a sample of 10000 muons with $\eta = 0$ generated with \texttt{acts-fatras} with the TrackML detector.}
\label{fig:pull}
\end{figure}

Transcription of the well-tested Combinatorial Kalman Filter (CKF) in ATLAS tracking software to ACTS design taking advantage of the high-performant ACTS \texttt{KalmanFilter} is planned. The ACTS \texttt{KalmanFilter} is designed to be independent of the EDM for tracks and allow for user-defined methods for filtering, smoothing, re-calibration of measurements using updated information from the track fitting and rejection of outlier measurements (measurements which are selected as incompatible with the track hypothesis). To record the track fitting results with the \texttt{KalmanFilter}, a flexible  \texttt{TrackState} class is used to contain various information relevant with fitting on a specific detector surface including measurements ( \texttt{uncalibrated} and  \texttt{calibrated}), fitted parameters (\texttt{predicted},  \texttt{filtered} and \texttt{smoothed}), fit quality, propagation length and a  \texttt{TrackStateFlag} as an interpreter of the \texttt{TrackState}, e.g. whether the  \texttt{TrackState} is an outlier or not. The \texttt{KalmanFilter} can be included in the Actors of a propagation call. To validate the performance of the  \texttt{KalmanFilter}, track fitting for truth trajectories generated with \texttt{acts-fatras} is performed. Figure~\ref{fig:trackeff} shows the almost $100\%$ fitting efficiency (the fraction of truth trajectories that are processed for smoothing) of \texttt{KalmanFilter} studied from a sample of 10000 muons with the TrackML detector simulated with \texttt{acts-fatras}. Figure~\ref{fig:pull} shows the distributions of the pulls of track parameters for 10000 muons generated with momentum direction $\eta = 0$. All the distributions are Gaussian with a width close to 1, which indicates that errors of track parameters in the \texttt{KalmanFilter} are correctly estimated. 

ACTS also includes a \texttt{GaussianSumFilter} prototype. The multi-component \texttt{MultiStepper} and \texttt{MultiMaterialInteractor} are implemented. Development of multi-component \texttt{MultiUpdater} and \texttt{MultiSmoother} is in progress. Tools for vertex reconstruction~\cite{vertex} including \texttt{IterativeVertexFinder} and \texttt{MultiAdaptiveVertexFitter} prototype have been implemented in ACTS as well.
     

\subsection{Alignment and calibration}
ACTS is designed for full parallel execution. Contextual data, i.e. detector alignment, calibration data, detector or magnetic field status are handled with context or payload objects that guarantee to provide access to the correct conditions in memory in a concurrent environment.
An \texttt{AlgorithmContext} which includes a \texttt{GeometryContext} object, a \texttt{CalibrationContext} object and a \texttt{MagneticFieldContext} object has been implemented to support on-the-fly alignment, calibration and magnetic field. The concept of \texttt{GeometryContext} has been demonstrated by running the propapation test with event-dependent alignment constants in flight. The concept of \texttt{CalibrationContext} using detector data has been successfully used in ATLAS. 
Implementation of a contexual calibration tool to apply additional calibration correction to the original measurement by importing the calibration tool used in ATLAS tracking software is planned. 

\section{Integration of ACTS to AthenaMT}
ACTS is designed to have no dependence on a particular experiment framework. A plugin mechanism is used in ACTS where necessary to allow interfacing experiment software. 

Work is on-going to integrate ACTS in AthenaMT. After building an ACTS \texttt{TrackingGeometry} for a specific ATLAS sub-detector, ACTS can run the test of propagation of particles through the geometry. The tracking geometry for the current ATLAS Inner Detector (ID) and Calorimeter in AthenaMT has been translated to ACTS \texttt{TrackingGeometry} and tested for particle prapagation. Implementation of ACTS \texttt{TrackingGeometry} for the ATLAS Muon Spectrometer is planned. 

The ACTS seed finder has been tested for single particle and the result have been validated with the ATLAS seed finder. It has also been tested for multi-threaded execution in AthenaMT. Track fitting with ACTS \texttt{KalmanFilter} and \texttt{GaussianSumFilter}, and vertex reconstruction with ACTS \texttt{IterativeVertexFinder} and \texttt{MultiAdaptiveVertexFitter} with trajectories through the ATLAS detectors are planned once prototypes of those fitters are complete and well validated.


\section{Summary}

The large increase in track multiplicity at future colliders needs tracking software of high performance with ability to exploit parallel architectures. The ACTS project aims to provide a framework-independent and detector-independent tracking
toolkit tested for strict thread-safety to support multi-threaded event processing. It is actively developed with collaboration across a range of experiments. 

Tracking components including tracking geometry, EDM, propagator, seed finder are well developed. The prototype of tracking deliverables for seed finding, track fitting and vertex reconstruction are available with performance in validation. Planned future developments will focus on implementation of tools for track finding and contexual re-calibration and alignment.

\section*{Acknowledgements}
This work was supported by the National Science Foundation under Cooperative Agreement OAC-1836650.

\end{document}